\documentclass[a4paper]{amsart}
\usepackage{amsmath,amsfonts,amssymb}
\usepackage{amscd}
\usepackage{amsthm}
\usepackage{latexsym}
\input xy
\xyoption{all}

 \newtheorem{thm}{Theorem}[section]
 \newtheorem{cor}[thm]{Corollary}
 \newtheorem{lem}[thm]{Lemma}
 \newtheorem{prop}[thm]{Proposition}
 \theoremstyle{definition}
 \newtheorem{defn}[thm]{Definition}
 \theoremstyle{remark}
 \newtheorem{rem}[thm]{Remark}
 \numberwithin{equation}{section}

\newlength{\defbaselineskip}
\newcommand{\setlinespacing}[1]%
           {\setlength{\baselineskip}{#1 \defbaselineskip}}

\begin{document}

\title [Universal Hopf algebra of renormalization ]
 {Universal Hopf algebra of renormalization and Hopf Algebras of Rooted Trees}

\author [Shojaei-Fard] {Ali Shojaei-Fard}
\address{Department of Mathematics, Shahid Beheshti University, 1983963113 Tehran, Iran}
\email{a\_shojaei@sbu.ac.ir}
\email{shojaeifa@yahoo.com}

\date{}

\dedicatory{}

\commby{}

\maketitle

\setlinespacing{1.11}

\begin{abstract}
\noindent In this paper we are going to find a rooted tree
representation from universal Hopf algebra of renormalization (in
Connes-Marcolli's approach in the study of renormalizable Quantum
Field Theories under the scheme minimal subtraction in dimensional
regularization). With attention to this new picture, interesting relations between
this specific Hopf algebra and some important Hopf algebras of
rooted trees and also Hopf algebra of (quasi-) symmetric functions
are obtained. And moreover a new interpretation from universal
singular frame, based on Hall rooted trees, is deduced such that it can be applied to the
physical information of a renormalizable theory such as
counterterms.
\end{abstract}

\section*{Introduction}

In modern physics the description of phenomena at the smallest
length scales with highest energies is done by Quantum Field Theory
(QFT) such that perturbative theory is a successful and useful approach to this powerful physical theory.
Renormalization plays the main part in this attitude such that its application on the perturbative expansions of divergent iterated Feynman integrals will deliver us renormalized values together with counterterms.

With arranging Feynman diagrams of a renormalizable
QFT into a Hopf algebra (such that its algebraic structures are induced
based on the existence of a recursive procedure for the elimination
of (sub)divergences from diagrams), Kreimer discovered a strong
mathematical interpretation for the perturbative
renormalization. With the concept of regularization, one can parametrize ultraviolet
divergences appearing in amplitudes to reduce them formally finite
together with a special subtraction of ill-defined expressions
(induced with physical principles). Moreover in the process of
regularization some non-physical parameters are created and this
fact will change the nature of Feynman rules to algebra morphisms
from the Hopf algebra (related to the given theory) to the
commutative algebra of Laurent series in dimensional regularization
such that in general this commutative algebra is characterized with
the given regularization method  \cite{EK1, F2}. It provides this fact that Feynman rules of the given theory can be
determined with some special characters of this Hopf algebra such that with
these Feynman rules, one can associate to each Feynman graph its
related amplitude. Therefore the lack of a practical mathematical basement
for this interesting physical technique in QFT is covered.
\cite{CK4, K6, K1, K10, K11, K4, K2}

Connes and Kreimer proved that perturbative renormalization can be
explained by a general mathematical procedure namely, extraction of
finite values based on the Riemann-Hilbert problem and in this way
they showed that one can obtain the important physical data of a
renormalizable QFT for instance renormalized values and counterterms
from the Birkhoff decomposition of characters of the related Hopf
algebra to the theory. In other words, they associated to each
theory an infinite dimensional Lie group and proved that in
dimensional regularization passing from unrenormalized to the
renormalized value is equivalent to the replacement of a given loop
(with values in the Lie group) with the value of the positive
component of its Birkhoff factorization at the critical integral
dimension $D$. In fact, in \cite{CK1, CK2} an algebraic
reconstruction from the Bogoliubov-Parasiuk-Hepp-Zimmermann (BPHZ)
method in renormalization is initiated.

Connes and Marcolli completed this mathematical machinery from
perturbative renormalization for renormalizable QFTs in the scheme
minimal subtraction in dimensional regularization underlying the
Riemann-Hilbert correspondence by giving a algebro-geometric
dictionary for the description of physical theories. In addition
they found a geometric manifestation from physical information of a given
theory by the classification of equisingular flat connections and
putting them in a category such that its objects give power to store
important parameters such as counterterms. By this way, each
counterterm is represented with the solution of a differential
system together with a special singularity, namely equisingularity.
In a general appearance, they introduced the universal category of flat
equisingular vector bundles $\mathcal{E}$ such that its universality
comes from this interesting notion that for each renormalizable
theory $\Phi$, one can put its related category of flat equisingular
connections $\mathcal{E}^{\Phi}$ as a subcategory in $\mathcal{E}$.
And because of the neutral Tannakian nature of this universal
category, one specific Hopf algebra will be characterized from the
process. That is universal Hopf algebra of renormalization $H_{U}$
such that by this special Hopf algebra and its affine group scheme,
renormalization groups and counterterms of all renormalizable physical
theories will have universal and canonical lifts. \cite{CM2, CM3,
CM1}

The combinatorial nature of the Hopf algebra of renormalization is
observed by its relation with the Connes-Kreimer Hopf algebra of
rooted trees \cite{K6, K10}. This Hopf algebra on rooted trees has
universal property with respect to the Hochschild cohomology theory
\cite{CK4, K11}. On the other hand, we know that the universal Hopf
algebra of renormalization has universal property with respect to
Hopf algebras (related to all renormalizable physical theories)
\cite{CM2}. These observations provide the notion of a closed relation
between the elements of $H_{U}$ and rooted trees.

In this work we are going to concentrate on this idea to obtain a new
picture from universal Hopf algebra of renormalization by rooted
trees. According to this goal, in the first section, we consider
Connes-Kreimer Hopf algebra of rooted trees. In the second part, the
basic definition of the universal Hopf algebra of renormalization is
studied. In the third part, we familiar with some interesting Hopf
algebra structures on rooted trees and also review an important
technique (i.e. operadic approach) for making an explicit rooted
tree interpretation from this universal element in the mathematical
treatment of perturbative renormalization. In section four, we make
rooted tree representations from $H_{U}$ at three levels: Hopf
algebra, affine group scheme and Lie algebra. Findings
interesting relations between this specific Hopf algebra and Hopf
algebra of (quai-) symmetric functions and also some important Hopf
algebras on rooted trees are immediate results from this rooted tree version. Finally in the last part of
this work, a rooted tree representation from universal singular
frame is given. Since this element maps to negative parts of the
Birkhoff decomposition of loops and provides universal counterterms,
therefore one can obtain a new image from counterterms.

\medskip

{\bf Acknowledgment.} The author acknowledges to support by
Hausdorff Research Institute for Mathematics (HIM) for the program
"Geometry-Physics" (May 2008 - August 2008), Max Planck Institute
for Mathematics (MPIM) as a member of IMPRS program (September 2008
- February 2009), Erwin Schrodinger International Institute for
Mathematical Physics (ESI) for the program "Number Theory and
Physics" (March 2009 - April 2009) and also partially research
fellowship from M.S.R.T (May 2008 - September 2008).

\section{Connes-Kreimer Hopf algebra of rooted trees}

Feynman diagrams together with the Feynman rules of a given
renormalizable QFT present all of the possible events in the theory
such that in general they contain (overlapping) sub-divergences. A
rooted tree representation from these diagrams will be applied for
at the first, their arranging into a Hopf algebraic structure and at
the second, describing more explicitly about the process of
renormalization. In the case of the overlapping divergences, at the
first step they should be reduced to a linear combination of
disjoint and nested divergences and next we have a sum of
rooted trees \cite{K1, KW1}. Depends on a fixed theory, all rooted
trees are equipped with specific decorations (which store
physical information such as (sub)-divergences) \cite{EK1, K6, K10,
K11}. Renormalization Hopf algebra of a given theory can be made based on
a decorated version of an interesting Hopf algebra structure on rooted
trees and in this part this Hopf algebra is studied.

A \textit{rooted tree} $t$ is an oriented, connected and simply
connected graph together with one distinguished vertex with no
incoming edge namely, root. A rooted tree $t$ with a given embedding
in the plane is called {\it planar rooted tree} and otherwise it is
called {\it (non-planar) rooted tree}.

Let $\textbf{T}$ be the set of all non-planar rooted trees and
$\mathbb{K} \textbf{T}$ be the vector space over a field
$\mathbb{K}$ (with characteristic zero) generated by $\textbf{T}$.
It is graded by the number of non-root vertices of rooted trees and
it means that
\begin{gather}
\bold{T_{n}}:= \{ t \in \bold{T}: |t|=n+1 \}, \   \   \
\bold{\mathbb{K}T}:= \bigoplus_{n \ge 0} \mathbb{K} \bold{T_{n}}.
\end{gather}
$H(\textbf{T}):=Sym(\mathbb{K} \textbf{T})$ is a graded free unital
commutative algebra containing $\mathbb{K} \textbf{T}$ such that the
empty tree is its unit. A monomial in rooted trees (that
commuting with each other) is called {\it forest}.

An \textit{admissible cut} $c$ of a rooted tree $t$ is a collection
of its edges with this condition that along any path from the root
to the other vertices, it meets at most one element of $c$. By
removing the elements of an admissible cut $c$ from a rooted tree
$t$, we will have a rooted tree $R_{c}(t)$ with the original root
and a forest $P_{c}(t)$ of rooted trees.

This concept determines a coproduct structure on $H(\textbf{T})$
given by
\begin{gather} \label{coproduct ck}
\Delta: H(\textbf{T}) \longrightarrow H(\textbf{T}) \otimes
H(\textbf{T}),  \   \  \Delta (t) = t \otimes \mathbb{I} +
\mathbb{I} \otimes t + \sum_{c} P_{c} (t) \otimes R_{c} (t)
\end{gather}
where the sum is over all possible non-trivial admissible cuts of
$t$. It is observed  that this coproduct can be rewritten in a
recursive way. Let $B^{+}:H(\textbf{T}) \longrightarrow
H(\textbf{T})$ be a linear operator that mapping a forest to a
rooted tree by connecting the roots of rooted trees in the forest to
a new root. $B^{+}$ is an isomorphism of graded vector spaces and
for the rooted tree $t=B^{+}(t_{1}...t_{n})$, we have

\begin{equation}
\Delta B^{+}(t_{1}...t_{n})=t \otimes \mathbb{I} + (id \otimes
B^{+}) \Delta (t_{1}...t_{n}).
\end{equation}
$\Delta$ is extended linearity to define it as an algebra
homomorphism.

In addition, one can define recursively an antipode on
$H(\textbf{T})$ given by
\begin{equation} \label{antip}
S(t)=-t- \sum_{c} S(P_{c}(t))R_{c}(t).
\end{equation}

And finally, we equip this space with the counit $\epsilon:
H(\textbf{T}) \longrightarrow \mathbb{K}$ given by
\begin{gather}
\epsilon(\mathbb{I})=1, \   \    \epsilon(t_{1}...t_{n})=0, \  \
t_{1}...t_{n}\neq \mathbb{I}.
\end{gather}

$H(\textbf{T})$ together with the coproduct (\ref{coproduct ck}) and
the antipode (\ref{antip}) is a finite type connected graded
commutative non-cocommutative Hopf algebra. It is called {\it
Connes-Kreimer Hopf algebra} and denoted by $H_{CK}$. \cite{CK4, K6,
K10}

Let $H(\textbf{T})^{\star}$ be the dual space that contains all
linear maps from $H(\textbf{T})$ to $\mathbb{K}$. A linear
map $f: H(\textbf{T}) \longrightarrow \mathbb{K}$ is called {\it
character}, if
\begin{gather}
f(t_{1}t_{2})=f(t_{1})f(t_{2}), \  \ f(\mathbb{I})=1.
\end{gather}
The set of all characters is denoted by \textbf{char H(\textbf{T})}.
A linear map $g: H(\textbf{T}) \longrightarrow \mathbb{K}$ is called
{\it derivation (infinitesimal character)}, if
\begin{gather}
g(t_{1}t_{2})=g(t_{1}) \epsilon(t_{2}) + \epsilon(t_{1})g(t_{2}).
\end{gather}
The set of all derivations is denoted by \textbf{$ \partial char
H(\textbf{T})$}.

One can equip the space $H(\textbf{T})^{\star}$ with the {\it
convolution product}
\begin{equation}
f * g (t):= m_{\mathbb{K}} (f \otimes g) \Delta(t).
\end{equation}
This product determines a group structure on \textbf{char
H(\textbf{T})} and a graded Lie algebra structure on \textbf{$
\partial char H(\textbf{T})$} such that there is a bijection map $\exp^{*}$ from \textbf{$ \partial char H(\textbf{T})$} to \textbf{char H(\textbf{T})}
(which plays an essential role in the presentation of components of
the Birkhoff decomposition of characters). \cite{EG2, EGK1, EGK2,
M2}

Finally one should notice to the {\it universal property} of this
Hopf algebra such that it is the result of the universal problem in
Hochschild cohomology.

\begin{thm} \label{1}
Let $\mathcal{C}$ be a category with objects $(H,L)$ consisting of a
commutative Hopf algebra $H$ and a Hochschild one cocycle $L:H
\longrightarrow H$. It means that for each $x \in H$,
$$\Delta L(x) = L(x) \otimes \mathbb{I} + (id \otimes L)\Delta(x).$$
And also Hopf algebra homomorphisms, that commute with cocycles, are
morphisms in this category. $(H_{CK},B^{+})$ is the universal
element in $\mathcal{C}$. In other words, for each object $(H,L)$
there exists a unique morphism of Hopf algebras $\phi: H_{CK}
\longrightarrow H$ such that $L \circ \phi = \phi \circ B^{+}$.
$H_{CK}$ is unique up to isomorphism. \cite{CK4, K6}
\end{thm}

\section {Universal Hopf algebra of renormalization}

The importance of this specific Hopf algebra in the mathematical
treatment of perturbative renormalization is studied in \cite{CM2, CM3, CM1} and here just we want to look at to the definition and
some basic properties of this Hopf algebra. Because we would like to
find a clear relation between the elements of this Hopf algebra and
rooted trees. The application of this relation in Connes-Marcolli's approach will be shown in the last section.

$H_{U}$ is a connected graded commutative non-cocommutative Hopf
algebra of finite type. It is the graded dual of the universal
enveloping algebra of the free graded Lie algebra $L_{\mathbb{U}}:=
\bold{F}(1,2,...)_{\bullet}$ generated by elements $e_{-n}$ of
degree $n>0$ (i.e. one generator in each degree). It is interesting
to know that as an algebra $H_{U}$ is isomorphic to the linear space
of noncommutative polynomials in variables $f_{n}$, $n \in
\mathbb{N}_{>0}$ with the shuffle product \cite{CM2, CM3}. It is a
gold key for us to find a relation between this Hopf algebra and
rooted trees but at the first we need some information about shuffle
structures.

Let $V$ be a vector space over the field $\mathbb{K}$ with the
tensor algebra $T(V)= \bigoplus_{n \ge 0} V^{\otimes n}$. Set

\begin{equation}
S(m,n)=\{ \sigma \in S_{m+n} : \sigma^{-1} (1) < ... < \sigma^{-1}
(m), \ \ \sigma^{-1} (m+1) < ... < \sigma^{-1} (m+n) \}.
\end{equation}
It is called the set of {\it $(m,n)-$shuffles}. For each $x=x_{1}
\otimes ... \otimes x_{m} \in V^{\otimes m}$, $y=y_{1} \otimes ...
\otimes y_{n} \in V^{\otimes n}$ and $\sigma \in S(m,n)$, define
\begin{equation}
\sigma (x \otimes y) = u_{\sigma (1)} \otimes u_{\sigma (2)} \otimes
... \otimes u_{\sigma (m+n)} \in V^{\otimes (m+n)}
\end{equation}
such that $u_{k} = x_{k}$ for $1\le k\le m$ and $u_{k}=y_{k-m}$ for
$m+1\le k\le m+n$. The {\it shuffle product} of $x,y$ is given by

\begin{equation} \label{shuffle}
x \star y := \sum_{\sigma \in S(m,n)} \sigma (x \otimes y).
\end{equation}
$(T(V),\star)$ is a unital commutative associative algebra and one
can define this product in a recursive procedure. There are some
extensions of this product such as {\it quasi-shuffles, mixable
shuffles}.

Let $A$ be a {\it locally finite set} (i.e. disjoint union of
finite sets $A_{n}, n \ge 1$). The elements of $A$ are {\it letters}
and monomials are called {\it words} such that the empty word is
denoted by $1$. Set $A^{-}:=A \cup \{0\}$. A locally finite set $A$
together with a Hoffman pairing $[.,.]$ on $A^{-}$ is called a {\it
Hoffman set}. Let $\mathbb{K}<A>$ be the graded noncommutative
polynomial algebra over the field $\mathbb{K}$. The {\it
quasi-shuffle product} $\star^{-}$ on $\mathbb{K}<A>$ is defined
recursively such that for any word $w$, $1 \star^{-} w = w \star^{-}
1 = w$ and also for words $w_{1}, w_{2}$ and letters $a, b$,
\begin{equation} \label{f1}
(aw_{1}) \star^{-} (bw_{2}) = a(w_{1} \star^{-} bw_{2}) + b(aw_{1}
\star^{-} w_{2})+[a,b](w_{1} \star^{-} w_{2}).
\end{equation}
$\mathbb{K}<A>$ together with the new product $\star^{-}$ is a
graded commutative algebra such that when $[.,.]=0$, it will be the
shuffle algebra $(T(V),\star)$ where $V$ is a vector space generated
by the set $A$.

\begin{thm} \label{2}

(i) There is a graded connected commutative non-cocommutative Hopf
algebra structure (of finite type) on $(\mathbb{K}<A>, \star^{-})$
($(T(V),\star)$) such that its coproduct is compatible with the
(quasi-)shuffle product.

(ii) There is an isomorphism (as a graded Hopf algebras) between
$(T(V),\star)$ and $(\mathbb{K}<A>, \star^{-})$.

(iii) There is a graded connected Hopf algebra structure (of finite
type) (comes from (quasi-)shuffle product) on the graded dual of
$\mathbb{K}<A>$.

(iv) We can extend the isomorphism in the second part to the graded
dual level. \cite{EG1, H1}

\end{thm}

\begin{proof}
The compatible Hopf algebra structure on the shuffle algebra of
noncommutative polynomials is given by the coproduct
\begin{equation}
\Delta (w)= \sum_{uv=w} u \otimes v
\end{equation}
and the counit
\begin{gather}
\epsilon (1)=1,  \  \  \  \epsilon (w)=0, \ w \neq 1.
\end{gather}

For a given Hoffman pairing $[.,.]$ and any finite sequence $S$ of
elements of the set $A$, with induction define $[S] \in A^{-}$ such
that for any $a \in A$, $[a]=a$ and $[a, S]=[a,[S]]$. Let $C(n)$ be
the set of compositions of $n$ and $C(n,k)$ be the set of
compositions of $n$ with length $k$. For each word $w=a_{1}...a_{n}$
and composition $I=(i_{1},...,i_{l})$, set
\begin{equation}
I[w] := [a_{1},...,a_{i_{1}}]
[a_{i_{1}+1},...,a_{i_{1}+i_{2}}]...[a_{i_{1}+...+i_{l-1}+1},...,
a_{n}].
\end{equation}
It means that compositions act on words. Now for any word
$w=a_{1}...a_{n}$, its antipode is given by
$$S(1)=1,$$
\begin{equation}
S(w)=- \sum_{k=0}^{n-1} S(a_{1}...a_{k}) \star^{-} a_{k+1}...a_{n}=
(-1)^{n} \sum_{I \in C(n)} I[a_{n}...a_{1}].
\end{equation}

To show the isomorphism between Hopf algebra structures (compatible
with the shuffle products) is done by the following maps
\begin{equation}
\tau(w)= \sum_{(i_{1},...,i_{l}) \in C(|w|)} \frac{1}{i_{1}! ...
i_{l}!} (i_{1},...,i_{l}) [w],
\end{equation}

\begin{equation}
\psi(w)=  \sum_{(i_{1},...,i_{l}) \in C(|w|)}
\frac{(-1)^{|w|-l}}{i_{1} ... i_{l}} (i_{1},...,i_{l}) [w].
\end{equation}
$\tau$ is an isomorphism of Hopf algebras and $\psi$ is its inverse.

The graded dual $\mathbb{K}<A>^{\star}$ has a basis consisting of
elements $v^{\star}$ (where $v$ is a word on $A$) with the following
pairing such that if $u=v$, then $(u,v^{\star})=1$ and if $u \neq
v$, then $(u,v^{\star})=0$. Its Hopf algebra structure is given by
the concatenation product
\begin{equation}
conc(u^{\star} \otimes v^{\star})=(uv)^{\star}
\end{equation}
and the coproduct
\begin{equation}
\delta (w^{\star}) = \sum_{u,v} (u \star^{-} v, w^{\star}) u^{\star}
\otimes v^{\star}.
\end{equation}

The map

\begin{equation}
\tau^{\star} (u^{\star}) = \sum_{n \ge 1} \sum_{[a_{1},...,a_{n}]=u}
\frac{1}{n!} (a_{1}...a_{n})^{\star},
\end{equation}
is an isomorphism in the dual level and its inverse is given by
\begin{equation}
\psi^{\star} (u^{\star})= \sum_{n \ge 1} \frac {(-1)^{n-1}}{n}
\sum_{[a_{1},...,a_{n}]=u} (a_{1}...a_{n})^{\star}.
\end{equation}
\end{proof}

Let $\mathcal{L}$ be a Lie algebra over the field $\mathbb{K}$.
There exists an associative algebra $\mathcal{L}_{0}$ over
$\mathbb{K}$ together with a Lie algebra homomorphism $\phi_{0}:
\mathcal{L} \longrightarrow \mathcal{L}_{0}$ such that for each
couple $(\mathcal{A}, \phi: \mathcal{L} \longrightarrow
\mathcal{A})$ of an algebra and a Lie algebra homomorphism, there is
a unique algebra homomorphism $\phi_{\mathcal{A}}:\mathcal{L}_{0}
\longrightarrow \mathcal{A}$ such that $\phi_{\mathcal{A}} \circ
\phi_{0}=\phi$. $\mathcal{L}_{0}$ is called {\it universal
enveloping algebra} of $\mathcal{L}$ and it is unique up to
isomorphism. The universal enveloping algebra $\mathcal{L}_{0}$ of
the free Lie algebra $\mathcal{L}(A)$ is a free associative algebra
on $A$ and $\phi_{0}$ is injective such that
$\phi_{0}(\mathcal{L}(A))$ will be the Lie subalgebra of
$\mathcal{L}_{0}$ generated by $\phi_{0} \circ i (A)$. \cite{R1}

The set of {\it Lie polynomials} in $\mathbb{K}<A>^{\star}$ is the
smallest sub-vector space of $\mathbb{K}<A>^{\star}$ containing the
set of generators $A^{\star}:=\{ a^{\star} : a \in A \}$ and closed
under the Lie bracket.

\begin{cor}  \label{4}
 The set of Lie polynomials in $\mathbb{K}<A>^{\star}$ forms a Lie
algebra. It is the free Lie algebra on $A^{\star}$ such that
$\mathbb{K}<A>^{\star}$ is its universal enveloping algebra.
\end{cor}

For the given locally finite set $A = \{ f_{n}:
n \in \mathbb{N}_{>0} \}$, let $V$ be its generated vector space. As an algebra, the universal
Hopf algebra $H_{U}$ is isomorphic to $(T(V),\star)$ and therefore its Hopf algebra structure is introduced by theorem \ref{2}.
At the Lie algebra level, we have to go to the dual structure. Result \ref{4} shows that
the set of all Lie polynomials in $H_{U}^{\star}$ is the free Lie
algebra generated by $\{f^{\star}_{n}\}_{n \in \mathbb{N}_{>0}}$
such that $H_{U}^{\star}$ is its universal enveloping algebra and on
the other hand, we know that $H_{U}^{\star}$ is identified by the
universal enveloping of the free graded Lie algebra $L_{\mathbb{U}}$
generated by $\{e_{-n}\}_{n \in \mathbb{N}_{>0}}$. Therefore it
makes sense that
\begin{equation} \label{order}
e_{-n} \longleftrightarrow f_{n}^{\star}.
\end{equation}

\section{Some interesting Hopf algebras of rooted trees}

Because of the importance of a toy model from Hopf algebra of Feynman graphs for practicers in theoretical physics,
rooted trees found an essential role in the study of QFT \cite{CK1, EK1}.
There are different kinds of Hopf algebra structures on rooted trees
and in this part with notice to the Connes-Kreimer renormalization, we
consider a group of Hopf algebras related to the Connes-Kreimer Hopf algebra and then relations among them will be discussed.

Define a noncommutative product $\bigcirc$ on $\bold{\mathbb{K}T}$.
Let $t,s$ be rooted trees such that  $t=B^{+}(t_{1}...t_{n})$ and
$|s|=m$. $t \bigcirc s$ is the sum of rooted trees given by
attaching each of $t_{i}$ to a vertex of $s$. One can define a
coproduct compatible with $\bigcirc$ on $\bold{\mathbb{K}T}$
given by
\begin{equation}
\Delta_{GL} B^{+}(t_{1},...t_{k}) = \sum_{I \cup J = \{1,2, ...,k\}}
B^{+}(t(I)) \otimes B^{+}(t(J)).
\end{equation}
$H_{GL}:=(\bold{\mathbb{K}T}, \bigcirc, \Delta_{GL})$ is a connected
graded noncommutative cocommutative Hopf algebra and it is called
{\it Grossman-Larson Hopf algebra}. $H_{GL}$ is the graded dual of
$H_{CK}$ and it is the universal enveloping algebra of its Lie
algebra of primitives. \cite{H3, M1, P1}

Let $\bold{P}$ be the set of all planar rooted trees and
$\bold{\mathbb{K}P}$ be its graded vector space. Tensor algebra
$T(\bold{\mathbb{K}P})$ is an algebra of ordered forests of planar
rooted trees and $B^{+}:T(\bold{\mathbb{K}P}) \longrightarrow
\bold{\mathbb{K}P}$ is an isomorphism of graded vector spaces. There
are two interesting Hopf algebra structures on $\bold{P}$.

One can represent planar rooted trees by the symbols $<, >$
together with some special rules. It is called {\it balanced bracket
representation or BBR}. Empty BBR is the representation of a tree
with just one vertex and in the BBR of a planar rooted tree of
degree $n$, each of the symbols $<$ and $>$ occur $n$ times and it
means that in reading from left to right, the count of $<$'s is
agree with the count of $>$'s.

A BBR $F$ is called {\it irreducible}, if $F=<G>$ for some BBR $G$
and otherwise it can be written by a juxtaposition $F_{1}...F_{k}$
of irreducible BBRs. These components correspond to the branches of
the root in the associated planar rooted tree.

Let $t,s$ be two planar rooted trees with BBR representations
$F_{t}, F_{s}$ such that $F_{t}=F^{1}_{t}...F^{k}_{t}$. Define a new
product $t \diamond s$ by a sum of planar rooted trees such that
their BBRs are given by shuffling the components of $F_{t}$ into the
$F_{s}$. Moreover the decomposition of elements into their
irreducible components determines a compatible coproduct
$\Delta_{\diamond}$ on $\bold{\mathbb{K}P}$. By these new operations
(given based on the balanced bracket representation), there is a
connected graded noncommutative Hopf algebra structure on
$\bold{\mathbb{K}P}$ and it is denoted by $H_{\bold{P}}:=
(\bold{\mathbb{K}P}, \diamond, \Delta_{\diamond})$. \cite{H2}

Another important Hopf algebra structure on planar rooted trees is
defined on the tensor algebra $T(\bold{\mathbb{K}P})$ together with
the given coproduct structure in (\ref{coproduct ck}) (on planar
rooted trees instead of rooted trees) such that it gives a graded
connected noncommutative Hopf algebra. It is called {\it Foissy Hopf
algebra} and denoted by $H_{F}$. $H_{F}$ is self-dual and isomorphic
to $H_{\bold{P}}$. \cite{H3, H2, H5}

Let $\mathbb{K}[[x_{1},x_{2},...]]$ be the ring of formal power
series. A formal series $f$ is called {\it symmetric
(quasi-symmetric)}, if the coefficients in $f$ of the monomials
$x^{i_{1}}_{n_{1}} ... x^{i_{k}}_{n_{k}}$ and $x^{i_{1}}_{1} ...
x^{i_{k}}_{k}$ equal for any sequence of distinct positive integers
$n_{1},...,n_{k}$ (for any increasing sequence $n_{1} < ... <
n_{k}$). In other words, we know that the symmetric groups
$\mathbb{S}_{n}$ act on $\mathbb{K}[[x_{1},x_{2},...]]$ by permuting
the variables and a symmetric function is invariant under these
actions and it means that after each permutation coefficients of its
monomials remain without any change. \cite{GKLLRT}

Let $SYM$ ($QSYM$) be the set of all symmetric (quasi-symmetric)
functions. It is easy to see that $SYM \subset QSYM$. As a vector
space, $QSYM$ is generated by the monomial quasi-symmetric functions
$M_{I}$ such that $I=(i_{1},...,i_{k})$ and $M_{I}:=\sum_{n_{1} <
n_{2} < ... < n_{k}} x^{i_{1}}_{n_{1}} ... x^{i_{k}}_{n_{k}}$ and if
we forget order in a composition, then we will reach to the
generators $m_{\lambda} := \sum_{\phi(I)= \lambda} M_{I}$ for $SYM$
(viewed as a vector space). \cite{H2, H5}

One can show that there is a graded connected commutative
cocommutative self-dual Hopf algebra structure on $SYM$ and also
there is a graded connected commutative non-cocommutative Hopf
algebra structure on $QSYM$ such that its graded dual is denoted by
$NSYM$. As an algebra, $NSYM$ is the noncommutative polynomials on
the variables $z_{n}$ of degree $n$. \cite{GKLLRT, H2, H5}

(Quasi-)symmetric functions play an interesting role to identify
some relations between this type of functions and various defined
Hopf algebras on (planar) rooted trees.

\begin{thm} \label{5}
There are following commutative diagrams of Hopf algebra
homomorphisms.  \cite{H2}

\begin{equation}
\begin{CD}
NSYM @>{\alpha_{1}}>> H_{F}\\
@V{\alpha_{3}}VV @V{\alpha_{2}}VV\\
SYM @>{\alpha_{4}}>> H_{CK}
\end{CD}
\end{equation}

\begin{equation}
\begin{CD}
SYM @<{\alpha_{4}^{\star}}<< H_{GL}\\
@V{\alpha_{3}^{\star}}VV @V{\alpha_{2}^{\star}}VV\\
QSYM @<{\alpha_{1}^{\star}}<< H_{\bold{P}}
\end{CD}
\end{equation}
\end{thm}

\begin{proof}
- $\alpha_{1}$ associates each variable $z_{n}$ to the ladder tree
$l_{n}$ of degree $n$.

- $\alpha_{2}$ maps each planar rooted tree to its corresponding
rooted tree without notice to the order in products.

- $\alpha_{3}$ sends each $z_{n}$ to the symmetric function
$m_{\underbrace{(1,...,1)}_{n}}$.

- $\alpha_{4}$ maps $m_{\underbrace{(1,...,1)}_{n}}$ to the ladder
tree $l_{n}$.

- For the composition $I=(i_{1},...,i_{k})$ one can define a planar
rooted tree $t_{I}:=B^{+}(l_{i_{1}},...,l_{i_{k}})$. For each planar
rooted tree $t$, if $t=t_{I}$, then define
$\alpha_{1}^{\star}(t):=M_{I}$ and otherwise
$\alpha_{1}^{\star}(t):=0$.

- For each rooted tree $t$, $\alpha_{2}^{\star}(t):=|sym(t)| \sum_{s
\in \alpha_{2}^{-1}(t)} s$.

- $\alpha_{3}^{\star}$ is the inclusion map.

- For the partition $J=(j_{1},...,j_{k})$, define a rooted tree
$t_{J}:=B^{+}(l_{j_{1}},...,l_{j_{k}})$. For each rooted tree $t$,
if $t=t_{J}$ (for some partition $J$), then define
$\alpha_{4}^{\star}(t):=|sym(t_{J})|m_{J}$ and otherwise
$\alpha_{4}^{\star}(t):=0$.

\end{proof}

{\it Zhao's homomorphism} provides another important group of
relations between rooted trees and (quasi-)symmetric functions. This
map is defined by

\begin{gather} \label{zhao}
Z: NSYM \longrightarrow H_{GL},  \  \  Z(z_{n}) = \epsilon_{n}
\end{gather}
such that rooted trees $\epsilon_{n}$ are given recursively by
$$\epsilon_{0}:= \bullet$$
\begin{equation}
\epsilon_{n}:=k_{1} \bigcirc \epsilon_{n-1} - k_{2} \bigcirc
\epsilon_{n-2} + ... + (-1)^{n-1} k_{n}
\end{equation}
where
\begin{equation}
k_{n}:= \sum_{|t|=n+1} \frac{t}{|sym(t)|} \in H_{GL}.
\end{equation}
$Z$ is an injective homomorphism of Hopf algebras and its dual can
be clarified uniquely. Define a linear map

\begin{equation}
A^{+}: QSYM \longrightarrow QSYM, \   \   M_{I} \longmapsto M_{I
\sqcup (1)}.
\end{equation}
For each ladder tree $l_{n}$ and monomial $u$ of rooted trees, the
surjective homomorphism $Z^{\star}: H_{CK} \longrightarrow QSYM $ is
given by

$$Z^{\star}(l_{n}):=M_{\underbrace{(1,...,1)}_{n}},$$
\begin{equation} \label{aplus}
Z^{\star} (B^{+}(u)):= A^{+} (Z^{\star}(u)).
\end{equation}
With the cocycle property of the map $A^{+}$, one can show that
$Z^{\star}$ is the unique homomorphism with respect to the relation
(\ref{aplus}) \cite{H5, H6}. In the next section, we will lift the
Zhao's homomorphism and its dual to the level of the  universal Hopf
algebra of renormalization.

Operadic approach to the poset theory provides an interesting source
of Hopf algebras namely, incidence Hopf algebras such that because
of the relation between Connes-Kreimer Hopf algebra and this class
of Hopf algebras, one can find the application of this part of
mathematics in the study of perturbative renormalization. In the
final part of this section this theory will be considered.

A partially ordered set ({\it poset}) is a set with a partial order
relation. A growing sequence of the elements of a poset is called
{\it chain}. A poset is {\it pure}, if for any $x \le y$ the maximal
chains between $x$ and $y$ have the same length. A bounded and pure
poset is called {\it graded poset}. For instance one can define a
graded partial order on the set $[n]=\{1,2,...,n\}$ by the
refinement of partitions and it is called {\it partition poset}.

A collection $\{P(n)\}_{n}$ of (right) $\mathbb{S}_{n}-$modules is
called {\it $\mathbb{S}-$module}. An {\it operad} $(P, \mu, \eta)$
is a monoid in the monoidal category
$\mathbb{S}-Mod:=(\mathbb{S}-Mod, \circ, \mathbb{I})$ of
$\mathbb{S}-$modules and it means that the composition morphism
$\mu: P \circ P \longrightarrow P$ is associative and the morphism
$\eta: \mathbb{I}\longrightarrow P$ is unit. This operad is called
{\it augmented}, if there exists a morphism of operads $\psi: P
\longrightarrow \mathbb{I}$ such that $\psi \circ \eta=id$.

A {\it $\mathbb{S}-$set} is a collection $\{P_{n}\}_{n}$ of sets
$P_{n}$ equipped with an action of the group $\mathbb{S}_{n}$. A
monoid $(P, \mu, \eta)$ in the monoidal category of
$\mathbb{S}-$sets is called a {\it set operad}. For each
$(x_{1},...,x_{t}) \in P_{i_{1}} \times ... \times P_{i_{t}}$, one
can define the map
$$\lambda_{(x_{1},...,x_{t})}: P_{t} \longrightarrow P_{i_{1}+...+i_{t}}$$
\begin{equation}
x \longmapsto \mu(x \circ (x_{1},...,x_{t})).
\end{equation}
A set opeard $P$ is called {\it basic}, if each
$\lambda_{(x_{1},...,x_{t})}$ is injective.

Partition posets associated to an operad is interesting for us. For
the set operad $(P, \mu, \eta)$ and each $n$, there is an action of
the group $\mathbb{S}_{n}$ on $P_{n}$. For the set $A$ with $n$
elements, let $\mathbb{A}$ be the set of ordered sequences of the
elements of $A$ such that each element appearing once. For each
element $x_{n} \times (a_{i_{1}},...,a_{i_{n}})$ in $P_{n} \times
\mathbb{A}$, its image under an element $\sigma$ of $\mathbb{S}_{n}$
is given by $\sigma(x_{n}) \times
(a_{\sigma^{-1}(i_{1})},...,a_{\sigma^{-1}(i_{n})})$. It is called
{\it diagonal action} and its orbit is denoted by $\overline{x_{n}
\times (a_{i_{1}},...,a_{i_{n}})}$. Let $\mathfrak{P}_{n}(A):=P_{n}
\times_{\mathbb{S}_{n}} \mathbb{A}$ be the set of all orbits under
this action. Set

\begin{equation}
P(A):=(\bigsqcup_{f:[n] \longrightarrow^{bijection} A} P_{n})_{\sim}
\end{equation}
where $(x_{n},f) \sim (\sigma(x_{n}),f \circ \sigma^{-1})$ is an
equivalence relation.  A {\it $P-$partition} of $[n]$ is a set of
components $B_{1},...,B_{t}$ such that each $B_{j}$ belongs to
$\mathfrak{P}_{i_{j}} (I_{j})$ where $i_{1}+...+i_{t}=n$ and
$\{I_{j}\}_{1 \le j \le t}$ is a partition of $[n]$. We can extend
maps $\lambda_{(x_{1},...,x_{t})}$ to $\lambda^{\sim}$ at the level
of $P(A)$ and it means that
$$ \lambda^{\sim}: P_{t} \times (\mathfrak{P}_{i_{1}}(I_{1}) \times ... \times \mathfrak{P}_{i_{t}}(I_{t})) \longrightarrow   \mathfrak{P}_{i_{1}+...+i_{t}}(A)$$
\begin{equation}
x \times (c_{1},...,c_{t}) \longmapsto \overline{\mu( x \circ
(x_{1},...,x_{t})) \times (a^{1}_{1},...,a^{t}_{i_{t}})}
\end{equation}
such that $\{I_{j}\}_{1 \le j \le t}$ is a partition of $A$ and each
$c_{r}$ is represented by $\overline{x_{r} \times
(a^{r}_{1},...,a^{r}_{i_{r}})}$ where $x_{r} \in P_{i_{r}}$,
$I_{r}=\{ a^{r}_{1},...,a^{r}_{i_{r}} \}$.

For the set operad $P$ and $P-$partitions $\mathfrak{B} =
\{B_{1},...,B_{r}\}$, $\mathfrak{C} = \{C_{1},...,C_{s}\}$ of $[n]$
such that $B_{k} \in \mathfrak{P}_{i_{k}} (I_{k})$ and $C_{l} \in
\mathfrak{P}_{j_{l}} (J_{l})$, we say that the $P-$partition
$\mathfrak{C}$ is larger than $\mathfrak{B}$, if for any $k \in
\{1,2,...,r\}$ there exists the subset $\{p_{1},...,p_{t}\} \subset
\{1,2,...,s\}$ such that $\{J_{p_{1}},...,J_{p_{t}}\}$ is a
partition of $I_{k}$ and if there exists an element $x_{t} \in
P_{t}$ such that $B_{k}=\lambda^{\sim} (x_{t} \times
(C_{p_{1}},...,C_{p_{t}}))$. This poset is called {\it operadic
partition poset} associated to the operad $P$ and denoted by
$\Pi_{P}([n])$. \cite{S1, V1}

One can extend the notion of this poset to each locally finite set
$A= \bigsqcup A_{n}$ such that in this case a $P-$partition of $[A]$
is a disjoint union (composition) of $P-$partitions of $[A_{n}]$s
and therefore the operadic partition poset associated to the operad
$P$ will be a composition of posets $\Pi_{P}([A_{n}])$ and denoted
by $\Pi_{P}([A])$.

A collection $(\mathfrak{p}_{i})_{i \in I}$ of posets is called {\it
good collection}, if each poset $\mathfrak{p}_{i}$ has a minimal
element $\bold{0}$ and a maximal element $\bold{1}$ (an interval)
and also for all $x \in \mathfrak{p}_{i} \ (i \in I)$, the interval
$[\bold{0},x]$ ($[x,\bold{1}]$) is isomorphic to a product of posets
$\prod_{j} \mathfrak{p}_{j}$ ($\prod_{k} \mathfrak{p}_{k}$).

For a given good collection $\mathcal{A}:= (\mathfrak{p}_{i})_{i \in
I}$, it is possible to make a new good collection $\mathcal{A}^{-}$
of all finite products $\prod_{i} \mathfrak{p}_{i}$ of elements such
that it is closed under products and closed under taking
subintervals. Let $[\mathcal{A}]$ ($[\mathcal{A}^{-}]$) be the set
of isomorphism classes of posets in $\mathcal{A}$
($\mathcal{A}^{-}$) such that elements in these sets denoted by
$[i], [j], ...$ and $H_{\mathcal{A}}$ be a vector space generated by
the elements $\{F_{[i]}\}_{[i] \in [\mathcal{A}^{-}]}$. It is
equipped with a commutative product (i.e. direct product of posets)
$F_{[i]}F_{[j]} = F_{[i \times j]}$ such that $F_{[e]}$ is the unit
(where $[e]$ is the isomorphism class of the singleton interval).
One should notice that as an algebra $H_{\mathcal{A}}$ may not be
free. By the concept of subinterval, there is a coproduct structure
on $H_{\mathcal{A}}$ given by

\begin{equation} \label{inc-co}
\Delta (F_{[i]}) = \sum_{x \in \mathfrak{p}_{i}} F_{[\bold{0},x]}
\otimes F_{[x,\bold{1}]}.
\end{equation}
With the coproduct (\ref{inc-co}), $H_{\mathcal{A}}$ has a
commutative Hopf algebra structure.

\begin{thm} \label{6}
Let $\Pi_{P}$ be a family of the operadic partition posets
associated to the set operad $P$. There is a good collection of
posets $(\mathfrak{p}_{i})$ (determined with $\Pi_{P}$). Its
associated Hopf algebra $H_{P}$ is called {\it incidence Hopf
algebra}. \cite{CL1, S1}
\end{thm}

One important note is that $H_{P}$ has a basis indexed by
isomorphism classes of intervals in the posets $\Pi_{P}(I)$ (for all
sets $I$) and this identification makes the sets $I$ disappear and
it means that the construction of this Hopf algebra is independent
of any label.

A rooted tree looks like a poset with a unique minimal element
(root) such that for any element $v$, the set of elements descending
$v$ forms a chain (i.e. the graph has no loop) and maximal elements
are called {\it leaves}. There is an interesting basic set operad on
rooted trees such that we will see that its incidence Hopf algebra
is isomorphic to the Connes-Kreimer Hopf algebra.

For the set $I$ with the partition $\{J_{i}\}_{i \ge 1 }$, suppose
$NAP(I)$ be the set of rooted trees with vertices labeled by $I$.
For $s_{i} \in NAP(J_{i})$ and $t \in NAP(I)$, we consider the
disjoint union of the rooted trees $s_{i}$ such that for each edge
of $t$ between $i_{1}, i_{2}$ in $I$, add an edge between the root
of $s_{i_{1}}$ and the root of $s_{i_{2}}$. The resulting graph is a
rooted tree labeled by $\bigsqcup_{i} J_{i}$ and its root is the
root of $s_{k}$ such that $k$ is the label of the root of $t$. It
gives us the composition $t ((s_{i})_{i \in I})$. In a more general
setting, it is observed that the operad $NAP$ plays the role of a
functor from the groupoid of sets to the category of sets. The
operadic partition poset $\Pi_{NAP} (I)$ is a set of forests of
$I-$labeled rooted trees such that a forest $x$ is covered by a
forest $y$, if $y$ is obtained from $x$ by grafting the root of one
component of $x$ to the root of another component of $x$. Or $x$ is
obtained from $y$ by removing an edge incident to the root of one
component of $y$.

Any interval in $\Pi_{NAP} (I)$ is a product of intervals of the
form $[\bold{0},t_{i}]$ such that $t_{i} \in NAP(J_{i})$. If
$t=B^{+}(t_{1}...t_{k})$, then the poset $[\bold{0},t]$ is
isomorphic to the product of the posets $[\bold{0}, B^{+}(t_{i})]$
for $i \in \{1,2,...,k\}$.

The incidence Hopf algebra $H_{NAP}$ is a free commutative algebra
on unlabeled rooted trees of root-valence $1$ such that elements
$F_{[t]}$ (where $t$ is a rooted tree) form a basis at the vector
space level. According to \ref{1} and the structure of $H_{NAP}$,
one can obtain the next important fact.

\begin{thm} \label{8}
$H_{NAP}$ is isomorphic to $H_{CK}$ by the unique Hopf algebra
isomorphism $\rho: F_{[B^{+} (t_{1}...t_{k})]} \longmapsto
t_{1}...t_{k}$. \cite{CL1}
\end{thm}

By this result, an operadic picture from Connes-Kreimer Hopf algebra
of rooted trees is given such that it will be applied to obtain an
operadic representation from the universal Hopf algebra of
renormalization.

\section{Rooted tree version of the universal Hopf algebra of renormalization}

A Hopf algebra structure is hidden in the process of perturbative
renormalization such that its description is possible with help of
the Hopf algebra $H_{CK}$ of rooted trees and in fact, rooted trees
(equipped with decorations related to a given theory) play the role
of a simplified model. On the other hand, the universal affine group
scheme $\mathbb{U}$ governs the structure of divergences of all
renormalizable theories and the universality of $H_{U}$ comes back
to its independency from all theories \cite{CM2, CM1}. In this
section, we want to provide an explicit interpretation from $H_{U}$
by rooted trees at three different levels Hopf algebra, Lie algebra
and affine group scheme. With the help of this new representation,
we will obtain interesting relations between Hopf algebra $H_{U}$
and other defined Hopf algebras in the previous parts.

There is a natural partial order $\preceq$ on the set of all rooted
trees $\bold{T}$. We say $t \preceq s$, if $t$ can be obtained from
$s$ by removing some non-root vertices and edges and it implies that
$|t| \le |s|$. From now let $\bold{T(A)}$ ($\bold{F(A)}$) be the set
of all rooted trees (forests) labeled by the set $A$.

For $a \in A$, $t_{1}, ..., t_{m} \in \bold{T(A)}$ such that
$u=t_{1}...t_{m} \in \bold{F(A)}$, $B^{+}_{a}(u)$ is a labeled
rooted tree of degree $|t_{1}|+...+|t_{m}|+1$ obtained by grafting
the roots of $t_{1}, ..., t_{m}$ to a new root labeled by $a$. It is
clear that $B^{+}_{a}(\mathbb{I})$ is a rooted tree with just one
labeled vertex.

For $t \in \bold{T(A)}$ and $u \in \bold{F(A)}, $ define a new
element $t \circ u$ such that it is a labeled rooted tree of degree
$|t|+|u|$ given by grafting the roots of labeled rooted trees in $u$
to the root of $t$.

\begin{rem} \label{9}

(i) $\circ$ is not associative.

(ii) $\forall t \in \bold{T(A)},$ $\forall u,v \in \bold{F(A)}:$ $(t
\circ u) \circ v = t \circ (uv) = (t \circ v) \circ u$.

(iii) $t_{1} \circ ... \circ t_{m} \circ u = t_{1} \circ (t_{2}
\circ ... \circ (t_{m} \circ u)), \  \ t^{\circ k}=t \circ ... \circ
t$, $k$ times.

(iv) For each $u \in \bold{F(A)}$, let $per(u)$ be the number of
different permutations of the vertices of a labeled partially
ordered set that representing $u$. Then
$$per(\mathbb{I})=1, \  \  per(B^{+}_{a}(u))=per(u).$$
And if $u=\prod^{m}_{j=1} (t_{j})^{i_{j}}$, then
$$per(u)= \prod^{m}_{j=1} i_{j}! per(t_{j})^{i_{j}}.$$

(v) The bilinearity extension of $\circ$ to the linear combinations
of labeled rooted trees (linear combinations of labeled forests) is
also possible.
\end{rem}

\begin{defn} \label{10}
A set $\bold{H(T(A))}$ of labeled rooted trees is called {\it Hall
set}, if it has following conditions:

- There is a total order relation $>$ on $\bold{H(T(A))}$.

- If $a \in A$, then $B^{+}_{a}(\mathbb{I}) \in \bold{H(T(A))}$.

- For $a \in A$, $u \in \bold{F(A)} - \{\mathbb{I}\}$ such that $u =
t^{\circ r_{1}}_{1} ... t^{\circ r_{m}}_{m}$, $t_{1},...,t_{m} \in
\bold{H(T(A))}$, $r_{1},...,r_{m} \ge 1$, $t_{1} > ... > t_{m}$,
$$ B^{+}_{a}(u) \in \bold{H(T(A))}  \Longleftrightarrow t_{m} > B^{+}_{a}(t^{\circ r_{1}}_{1} ... t^{\circ r_{m-1}}_{m-1}) \in \bold{H(T(A))}. $$

- If $t = B^{+}_{a}(t^{\circ r_{1}}_{1} ... t^{\circ r_{m}}_{m}) \in
\bold{H(T(A))}$  such that $t_{1},...,t_{m} \in \bold{H(T(A))}$,
$r_{1},...,r_{m} \ge 1$, $a \in A$, then for each $j=1,...,m$,
$t_{j}
> t$.

\end{defn}
From definition, for each $t \in \bold{H(T(A))}$, $r \ge 1$ and $a
\in A$, it is easy to see that

\begin{equation}
B^{+}_{a}(t^{\circ r}) \in \bold{H(T(A))} \Longleftrightarrow t >
B^{+}_{a}(\mathbb{I}).
\end{equation}
For the Hall set $\bold{H(T(A))}$, the set of its forests is given
by
\begin{equation}
\bold{H(F(A))}:=\{\mathbb{I}\} \cup \{t^{r_{1}}_{1} ...
t^{r_{m}}_{m}: r_{1},...,r_{m} \ge 1, t_{1},...,t_{m} \in
\bold{H(T(A))}, t_{i} \neq t_{j} (i \neq j) \}.
\end{equation}
The relation between the elements of $\bold{H(F(A))}$ and rooted
trees is clarified  by the map
$$\xi: \bold{H(F(A))} - \{\mathbb{I}\} \longrightarrow \bold{T(A)}$$
\begin{equation}
t^{\circ r_{1}}_{1} ... t^{\circ r_{m}}_{m} \longrightarrow t^{\circ
r_{1}}_{1} \circ (t^{\circ r_{2}}_{2}...t^{\circ r_{m}}_{m}).
\end{equation}
$\xi$ is injective and its image is the set $\{B^{+}_{a}(u) \in
\bold{T(A)}: u \in  \bold{H(F(A))}, a \in A \}$. Hall trees and Hall
forests have no symmetry. There is a one to one correspondence
between a Hall set of $A-$labeled rooted trees and a Hall set of
words on $A$. \cite{R1}

For $t \in \bold{H(T(A))}$, there is a {\it standard decomposition}
$(t^{1},t^{2}) \in \bold{H(T(A))} \times \bold{H(T(A))}$ such that

- If $|t|=1$, the decomposition is $t^{1}=t,$ $t^{2}=\mathbb{I}$,

- And if $t= B^{+}_{a}(t^{\circ r_{1}}_{1} ... t^{\circ r_{m}}_{m})$
such that $r_{1},...,r_{m} \ge 1,$ $t_{1},...,t_{m} \in
\bold{H(T(A))}:$  $t_{1} > ... > t_{m},$ $a \in A$, the
decomposition is given by
\begin{gather}
t^{1}= B^{+}_{a} (t^{\circ r_{1}}_{1} ... t^{\circ
r_{m-1}}_{m-1}t^{\circ r_{m}-1}_{m}), \  \ t^{2}=t_{m}.
\end{gather}

- For a Hall forest $u \in \bold{H(F(A))} - \bold{H(T(A))}$ such
that $u=t^{\circ r_{1}}_{1} ... t^{\circ r_{m}}_{m}$, $t_{1} > ... >
t_{m}$, the decomposition is given by $(u^{1},u^{2}) \in
\bold{H(F(A))} \times \bold{H(T(A))}$ where
\begin{gather}
u^{1}=t^{\circ r_{1}}_{1} ... t^{\circ r_{m-1}}_{m-1}t^{\circ
r_{m}-1}_{m}, \  \  u^{2}=t_{m}.
\end{gather}

For a given map that associates to each word $w$ on $A$ a scalar
$\alpha_{w} \in \mathbb{K}$, define a map $\alpha: \bold{F(A)}
\longrightarrow \mathbb{K}$ such that

- $\mathbb{I} \longmapsto \alpha_{1}$.

- For each $u \in \bold{F(A)}-\{\mathbb{I}\}$, there is a labeled
partially ordered set $(\mathfrak{u}(A), \ge)$ that represents the
forest $u$ such that vertices $x_{1},...,x_{n},...$ of this poset
are labeled by $l(x_{i})=a_{i} \in A$ ($1 \le i$). Let
$>_{\mathfrak{u}(A)}$ be a total order relation on the set of
vertices $\mathfrak{u}(A)$ such that it is an extension of the
partial order relation $\ge$ on $\mathfrak{u}(A)$. For each ordered
sequence $x_{i_{1}} >_{\mathfrak{u}(A)} ...
>_{\mathfrak{u}(A)} x_{i_{n}}$ in $\mathfrak{u}(A)$, its
corresponding word $a_{i_{1}}...a_{i_{n}}$ is denoted by
$w(>_{\mathfrak{u}(A)})$. Set
\begin{equation} \label{f2}
\alpha (u) := \sum_{>_{\mathfrak{u}(A)}}
\alpha_{w(>_{\mathfrak{u}(A)})},
\end{equation}
where the sum is over all total order relations
$>_{\mathfrak{u}(A)}$ (i.e. extensions of the main partial order
relation $\ge$) on the set of vertices of $\mathfrak{u}(A)$.

One can define another map
\begin{gather} \label{11}
\pi: \mathbb{K}[\textbf{T(A)}] \longrightarrow
(\mathbb{K}<A>,\star^{-}),  \    \     \ \pi (u):=
\sum_{>_{\mathfrak{u}(A)}} w(>_{\mathfrak{u}(A)})
\end{gather}
such that for each $u, v \in \textbf{F(A)}$ and
$a \in A$, it is observed that
\begin{gather} \label{12}
\pi (B^{+}_{a}(u))=\pi(u)a,  \ \  \pi (uv)=\pi(u) \star^{-} \pi(v),
\ \  \alpha(u)= \widehat{\alpha}(\pi(u))
\end{gather}
where $\widehat{\alpha}: (\mathbb{K}<A>,\star^{-}) \longrightarrow
\mathbb{K},$ $\widehat{\alpha} (w)= \alpha_{w}$ is a
$\mathbb{K}-$linear map.

There is a canonical map $f$ on Hall rooted trees defined by
$f(a)=a$, if $a \in A$ and $f(t)=f(t^{1})f(t^{2})$, if $t$ is of
degree $\ge 2$ with the standard decomposition $t=(t^{1},t^{2})$.
The function $f$ is called {\it foliage} and for each Hall tree $t$,
its degree $|f(t)|$ is the number of leaves of $t$. The foliage of a
Hall tree is called {\it Hall word} and for each word $w$ on $A$,
there is a unique factorization $w=f(t_{1})...f(t_{n})$ such that
$t_{i} \in \bold{H(T(A))}$ and $t_{1} > ... > t_{n}$ and also one
can show that Hall sets of labeled rooted trees can be reconstructed
recursively from an arbitrary Hall set of words on $A$. \cite{R1}

Let $A$ be a totally ordered set. The {\it alphabetical ordering}
gives a total order on the set of words on $A$ such that for any
nonempty word $v$, put $u < uv$ and also for letters $a < b$ and
words $w_{1}, w_{2}, w_{3}$, put $w_{1}aw_{2} < w_{1}bw_{3}$. A
non-trivial word $w$ is called {\it Lyndon}, if for any non-trivial
factorization $w=uv$, we have $w<v$. The set of Lyndon words,
ordered alphabetically, is a Hall set. We want to show that the
consideration of its corresponding Hall set of labeled rooted trees
is interested.

\begin{thm} \label{shuffle-order}
Let $A$ be a locally finite set with a total order relation on it.
The (quasi-)shuffle algebra $(\mathbb{K}<A>, \star^{-})$ is the free
polynomial algebra on the Lyndon words. \cite{H1}
\end{thm}

We know that the universal Hopf algebra of renormalization (as an
algebra) is given by the shuffle product on the linear space of
noncommutative polynomials with variables $f_{n}$ ($n \in
\mathbb{N}$). With notice to the correspondence (\ref{order}), one
can define a natural total order relation (depended on degrees of
the generators $e_{-n},$ $(n \in \mathbb{N})$ of the free Lie
algebra $L_{\mathbb{U}}$) on the set $A = \{ f_{n}: n \in
\mathbb{N}_{>0} \}$. It is given by
\begin{equation} \label{totalorder}
f_{n} < f_{m} \Longleftrightarrow n > m.
\end{equation}
Now \ref{shuffle-order} shows that $H_{U}$ (as an algebra) should be
free polynomial algebra of the Lyndon words on the set $A$. Consider
the Hall set of these Lyndon words (ordered alphabetically) such
that its corresponding Hall set of labeled rooted trees is denoted
by $\bold{H(T(A))}_{U}$. It makes sense to say that Lyndon words
play the role of a bridge between rooted trees and $H_{U}$.

Let us consider free commutative algebra $\mathbb{K}[\textbf{T(A)}]$
such that the set $\{t^{r_{1}}_{1}...t^{r_{m}}_{m}: t_{1},...,t_{m}
\in \textbf{T(A)}\}$ is a $\mathbb{K}-$basis (as a graded vector
space) where each expression $t^{r_{1}}_{1}...t^{r_{m}}_{m}$ is a
forest. For the forest $u$ with the associated partial order set
$(\mathfrak{u}(A), \ge)$, its coproduct is given by

\begin{equation} \label{13}
\Delta (u)= \sum_{(\mathfrak{v}(A),\mathfrak{w}(A)) \in
R(\mathfrak{u}(A))} v \otimes w
\end{equation}
such that labeled forests $v,w$ are represented by labeled partially
ordered subsets $\mathfrak{v}(A),\mathfrak{w}(A)$ of
$\mathfrak{u}(A)$ together with the following properties:

- The set of vertices in $\mathfrak{u}(A)$ is the disjoint union of
the set of vertices $\mathfrak{v}(A)$ and $\mathfrak{w}(A)$,

- For each $x,y \in \mathfrak{u}(A)$ such that $x \ge y$, if $x \in
\mathfrak{w}(A)$ then $y \in \mathfrak{w}(A)$.

By (\ref{13}), there is a connected graded commutative Hopf algebra
structure on $\mathbb{K}[\textbf{T(A)}]$ such that the product in
the dual space $\mathbb{K}[\textbf{T(A)}]^{\star}$ corresponds to
the given coproduct (\ref{13}) namely, dual to the coalgebra
structure and it means that for each $\alpha, \beta \in
\mathbb{K}[\textbf{T(A)}]^{\star}$ and each forest $u$,
\begin{equation}
<\alpha \beta, u> = <\alpha \otimes \beta, \Delta(u)>.
\end{equation}
One can show that $H_{GL}$ (labeled by the set $A$) and
$\mathbb{K}[\textbf{T(A)}]$ are graded dual to each other and
moreover with respect to the coproduct structure (\ref{13}) and the
operator $B^{+}_{a}$, this Hopf algebra has the universal property.

\begin{thm} \label{14}
Let $H$ be a commutative Hopf algebra over the field $\mathbb{K}$
and $\{L_{a}: H \longrightarrow H \}$ be a family of
$\mathbb{K}-$linear maps such that $\cup_{a \in A} Im L_{a} \subset
ker \epsilon_{H}$ and $\Delta_{H}L_{a}(c)=L_{a}(c) \otimes
\mathbb{I}_{H} + (id_{H} \otimes L_{a})\Delta_{H}(c)$. Then there
exists a unique Hopf algebra homomorphism $\psi:
\mathbb{K}[\textbf{T(A)}] \longrightarrow H$ such that for each $u
\in \mathbb{K}[\textbf{T(A)}]$ and $a \in A$,
$\psi(B^{+}_{a}(u))=L_{a}(\psi(u))$. \cite{M1}
\end{thm}

If we look at to \ref{1} (universal property of the Connes-Kreimer
Hopf algebra), then it is easy to understand that \ref{14} is a
poset version of this universal object and it means that $H_{CK}$
(labeled by the set $A$) is isomorphic to
$\mathbb{K}[\textbf{T(A)}]$.

Now one can observe that the $\mathbb{K}-$linear map $\pi$ is a Hopf
algebra homomorphism and for each $a_{1},...,a_{m} \in A$, we have
\begin{equation}
\pi(B^{+}_{a_{m}}...B^{+}_{a_{2}}(a_{1})) = a_{1}...a_{m}.
\end{equation}
It provides a bijection between the set of non-empty words and the
set of labeled rooted trees without side-branchings. This shows that
$\pi$ is an epimorphism and for  each $\widehat{\alpha},
\widehat{\beta} \in \mathbb{K}<A>^{\star}$, $u \in
\mathbb{K}[\textbf{T(A)}]$, we have
\begin{equation}
<\widehat{\alpha}\widehat{\beta}, \pi(u)>=<\alpha \beta, u>
\end{equation}
such that $\alpha, \beta \in \mathbb{K}[\textbf{T(A)}]^{\star}$. On
the other hand, $I := Ker \pi$ is a Hopf ideal in
$\mathbb{K}[\textbf{T(A)}]$ and there is an explicit picture from
its generators given by

\begin{equation} \label{f6}
I = Ker \pi =
\end{equation}

$<\{ \prod^{m}_{i=1} t_{i} - \sum^{m}_{i=1} t_{i} \circ \prod_{j
\neq i} t_{j}: m>1, t_{1},...,t_{m} \in \textbf{T(A)} \}> = < \{ t
\circ z + z \circ t - tz :t,z \in  \textbf{T(A)} \} \cup \{ s \circ
t \circ z + s \circ z \circ t - s \circ (tz): t,z,s \in
\textbf{T(A)} \}> = <\{ t \circ z + z \circ t -tz : t,z \in
\textbf{T(A)} \} \cup \{ s \circ (tz) + z \circ (ts) + t \circ (sz)
- tzs: t,z,s \in \textbf{T(A)}\}>.$

\begin{thm} \label{15}
The (quasi-)shuffle Hopf algebra $(\mathbb{K}<A>,\star^{-})$ is
isomorphic to the quotient Hopf algebra
$\frac{\mathbb{K}[\textbf{T(A)}]}{I}$. As an $\mathbb{K}-$algebra,
$\frac{\mathbb{K}[\textbf{T(A)}]}{I}$ is freely generated by the set
$\{t + I: t \in \bold{H(T(A))}\}$.  \cite{M1}
\end{thm}

According to the above theorem and also the given operadic picture
from Connes-Kreimer Hopf algebra in the previous section, we can
reach to the following result.

\begin{cor} \label{16}
(i) For each locally finite set $A$ together with a total order
relation, there exist Hopf ideals $J_{1}, J_{2}$ such that
$$(\mathbb{K}<A>,\star^{-}) \cong \frac{\mathbb{K}[\textbf{T(A)}]}{I} \cong \frac{H_{CK}(A)}{J_{1}} \cong \frac{H_{NAP}(A)}{J_{2}}.$$

(ii) Universal Hopf algebra of renormalization is isomorphic to a
quotient of the (labeled) incidence Hopf algebra with respect to the
basic set operad NAP.

\end{cor}

\begin{proof}
By \ref{8}, \ref{15}, it is enough to set $J_{1}:=I$ and
$J_{2}:=\rho^{-1}{J_{1}}$. \ref{15} is a representation of $H_{U}$
by rooted trees. It is enough to replace the set $A$ with the
variables $f_{n}$ such that the identified Lyndon words (with the
shuffle structure of $H_{U}$) gives us the Hall set
$\bold{H(T(A))}_{U}$.
\end{proof}

One can obtain interesting relations between the introduced Hopf
algebras in the previous parts and $H_{U}$.

\begin{prop} \label{diag}
We have following commutative diagrams of Hopf algebra
homomorphisms.

\begin{equation}
\begin{CD}
NSYM @>{\beta_{1}}>> H_{F}\\
@V{\beta_{3}}VV @V{\beta_{2}}VV\\
SYM @>{\beta_{4}}>> H_{U}
\end{CD}
\end{equation}

\begin{equation}
\begin{CD}
SYM @<{\beta_{4}^{\star}}<< U(L_{\mathbb{U}})\\
@V{\beta_{3}^{\star}}VV @V{\beta_{2}^{\star}}VV\\
QSYM @<{\beta_{1}^{\star}}<< H_{\bold{P}}
\end{CD}
\end{equation}
\end{prop}

\begin{proof}

For each (planar) rooted tree $t$, one can put different labels
(with elements of the locally finite set $A=\{f_{n}\}_{n}$) on its
vertices. Let $[t]$ be the class of all different possible Hall
(planar) rooted trees with respect to $t$ in $\bold{H(T(A))}_{U}$
and for the forest $u$, let $[u]$ be the class of all different Hall
forests associated to $u$ in $\bold{H(F(A))}_{U}$. With notice to
the diagrams in \ref{5} and with help of the given results in
\ref{15} and \ref{16}, define

- $\beta_{1}:=\alpha_{1}$,

- $\beta_{3}:=\alpha_{3}$,

- For each forest $u$ of planar rooted trees, $\beta_{2}(u):=
\sum_{v \in [u]} \pi(v)$,

- For each symmetric function $m_{\underbrace{(1,...,1)}_{n}}$,
$\beta_{4}(m_{\underbrace{(1,...,1)}_{n}}):= \sum_{v \in [l_{n}]}
\pi(v)$,

- $\beta^{\star}_{1}:=\alpha^{\star}_{1}$,

- $\beta^{\star}_{3}:=\alpha^{\star}_{3}$,

- For each generator $e_{-n}$, $ \beta^{\star}_{4}(e_{-n}):=
\alpha^{\star}_{4}(l_{n})$,

- For each generator $e_{-n}$, $ \beta^{\star}_{2}(e_{-n}):=
\alpha^{\star}_{2}(l_{n})$.

\end{proof}

Now we want to lift the Zhao's homomorphism (\ref{zhao}) and its
dual to the level of  the universal Hopf algebra of renormalization.
We know that $\pi$ is a surjective homomorphism from $H_{CK}(A)$ to
$H_{U}$ and on the other hand, $Z^{\star}$ gives us a unique
surjective map from $H_{CK}$ to $QSYM$ with the property
(\ref{aplus}). For a word $w$ with length $n$ in $H_{U}$, there
exists labeled ladder tree $l^{w}_{n}$ of degree $n$ in $H_{CK}(A)$
such that $\pi(l^{w}_{n})=w$. Define a new map $Z_{u}: H_{U}
\longrightarrow QSYM$ such that it maps each element $w \in H_{U}$
to
\begin{equation} \label{exten-zhao1}
Z_{u}(w):=Z^{\star}(l_{n}).
\end{equation}
It is observed that $Z_{u}$ is a unique homomorphism of Hopf
algebras. With the help of \ref{5} and \ref{diag}, one can define
homomorphisms $\theta_{1}:= \beta_{4} \circ \beta_{3} = \beta_{2}
\circ \beta_{1}$ from $NSYM$ to $H_{U}$ and $\theta_{2}:= \beta_{4}
\circ \alpha_{4}^{\star} = \beta_{2} \circ \alpha^{\star}_{2}$ from
$H_{GL}$ to $H_{U}$. On the other hand, the surjective morphism
$\pi$ determines a new homomorphism $\rho$ from $H_{CK}$ to $H_{U}$
such that for each unlabeled forest $u$ in $H_{CK}$, it is defined
by
\begin{equation}
\rho(u):= \sum_{v \in [u]} \pi(v).
\end{equation}
With notice to the homomorphisms (\ref{zhao}) and
(\ref{exten-zhao1}), the following commutative diagram can be
obtained.

\begin{equation} \label{hex1}
\begin{split}
\xymatrixcolsep{3pc} \xymatrix{
& SYM  \ar[ld]_{\alpha_{4}} \ar[dd]_{\beta_{4}} \\
H_{CK}  \ar[rd]^{\rho}  & & H_{GL}  \ar[ul]_{\alpha_{4}^{\star}} \ar[ld]^{\theta_{2}}\\
& H_{U} \ar[ld]_{Z_{u}}   &\\
QSYM   & & NSYM  \ar[ld]^{\alpha_{3}} \ar[lu]^{\theta_{1}}\\
& SYM \ar[lu]^{Z_{u} \circ \beta_{4}} \ar[uu]^{\beta_{4}} }
\end{split}
\end{equation}
The dual of the maps $Z_{u}$ and $\rho$ are given by
$$Z_{u}^{\star}: NSYM \longrightarrow U(L_{\mathbb{U}})$$
\begin{equation} \label{dualunzh}
Z_{u}^{\star}(z_{n}):=e_{-n},
\end{equation}
$$\rho^{\star}: U(L_{\mathbb{U}}) \longrightarrow H_{GL}$$
\begin{equation}
\rho^{\star}(e_{-n}):= l_{n}.
\end{equation}
With the help of these maps and with notice to \ref{diag} and the
given diagrams in \ref{5}, one can obtain a dual version of the
diagram (\ref{hex1}) given by
\begin{equation} \label{hex2}
\begin{split}
\xymatrixcolsep{3pc} \xymatrix{
& SYM  \ar[rd]^{\alpha_{4}} \\
H_{GL} \ar[ur]^{\alpha_{4}^{\star}} & & H_{CK} \\
& U(L_{\mathbb{U}}) \ar[uu]_{\beta_{4}^{\star}}
\ar[dd]^{\beta_{4}^{\star}}
\ar[lu]_{\rho^{\star}} \ar[rd]_{\theta_{1}^{\star}} \ar[ru]_{\theta_{2}^{\star}} &\\
NSYM \ar[ru]_{Z_{u}^{\star}} \ar[rd]_{(Z_{u} \circ \beta_{4})^{\star}}  & & QSYM \\
& SYM \ar[ru]_{\alpha_{3}^{\star}} }
\end{split}
\end{equation}

Graded dual relation for the pair $(H_{CK}(A),H_{GL}(A))$ and also
the pair $(H_{U},U(L_{\mathbb{U}}))$ provide another homomorphisms
from $H_{U}$ to $H_{GL}(A)$ and also from $H_{CK}(A)$ to
$U(L_{\mathbb{U}})$.

\begin{lem} \label{relation-dual}
Let $H_{1}$ and $H_{2}$ be graded connected locally finite Hopf
algebras which admit inner products $(.,.)_{1}$ and $(.,.)_{2}$,
respectively. If they are dual to each other, then there is a linear
map $\lambda: H_{1} \longrightarrow H_{2}$ such that

- $\lambda$ preserves degree,

- For each $h_{1}, h_{2} \in H_{1}:$
$(h_{1},h_{2})_{1}=(\lambda(h_{1}),\lambda(h_{2}))_{2},$

- For each $h_{1}, h_{2},  h_{3} \in H_{1}:$
$$(h_{1}h_{2},h_{3})_{1}=(\lambda(h_{1}) \otimes \lambda(h_{2}), \Delta_{2}(\lambda(h_{3})))_{2},$$
$$(h_{1} \otimes h_{2}, \Delta_{1}(h_{3}))_{1} = (\lambda(h_{1})\lambda(h_{2}), \lambda(h_{3}))_{2}.$$
This linear map determines an isomorphism $\tau: H_{2}
\longrightarrow H_{1}^{\star}$ such that for each $h_{1} \in H_{1}$
and $h_{2} \in H_{2}$, it is defined by
$$<\tau(h_{2}),h_{1}>:=(h_{2},\lambda(h_{1}))_{2}.$$ \cite{H2}
\end{lem}

Since $H_{CK}(A)$ and $H_{GL}(A)$ are graded dual to each other,
therefore by \ref{relation-dual} one can find a linear map
$\lambda:H_{CK}(A) \longrightarrow H_{GL}(A)$ with the mentioned
properties such that it defines an isomorphism $\tau_{1}$ from
$H_{GL}(A)$ to $H_{CK}(A)^{\star}$ given by

\begin{equation}
<\tau_{1}(t),s>:=(t, \lambda(s))
\end{equation}
where for rooted trees $t_{1},t_{2}$, if $t_{1}=t_{2}$ then
$(t_{1},t_{2})=|sym(t_{1})|$ and otherwise it will be $0$. For each
word $w$ with length $n$ in $H_{U}$, there is a labeled ladder tree
$l^{w}_{n}$ in $H_{CK}(A)$ such that $\pi(l^{w}_{n})=w$. Now one can
define a new homomorphism $F$ given by
$$F: H_{U} \longrightarrow H_{GL}(A)$$
\begin{equation} \label{alpha}
F(w):= \tau_{1}^{-1}((l^{w}_{n})^{\star}).
\end{equation}

Since $H_{U}$ and $U(L_{\mathbb{U}})$ are graded dual to each other,
therefore with the help of \ref{2} and \ref{relation-dual} one can
obtain a linear map $\theta: H_{U}  \longrightarrow
U(L_{\mathbb{U}})$ with the mentioned properties and after that  an
isomorphism $\tau_{2}$ from $U(L_{\mathbb{U}})$ to $H_{U}^{\star}$
will be determined. For each element $x \in U(L_{\mathbb{U}})$ and
word $w \in H_{U}$, it is given by
\begin{equation}
<\tau_{2}(x),w>:=(x, \theta(w))
\end{equation}
such that $(.,.)$ is the natural pairing on $U(L_{\mathbb{U}})$. For
a labeled forest $u$, $\pi(u)$ is an element in $H_{U}$. By the
natural pairing (given in \ref{2}), the dual of $F$ is identified by
$$ F^{\star}: H_{CK}(A) \longrightarrow  U(L_{\mathbb{U}})$$
\begin{equation} \label{alpha-dual}
F^{\star}(u):= \tau_{2}^{-1}((\pi(u))^{\star}).
\end{equation}

There is an interesting strategy to find a rooted tree
representation from the universal affine group scheme $\mathbb{U}$
by the theory of construction of a group from an operad. Let $P$ be
an augmented set operad and $\mathbb{K}A_{P}= \bigoplus_{n}
\mathbb{K} (P_{n})_{\mathbb{S}_{n}}$ be the direct sum of its
related coinvariant spaces with the completion
$\widehat{\mathbb{K}A_{P}}= \prod_{n} \mathbb{K}
(P_{n})_{\mathbb{S}_{n}}$. There is an associative monoid structure
on $\widehat{\mathbb{K}A_{P}}$. Let $\mathcal{G}_{P}$ be the set of
all elements of $\widehat{\mathbb{K}A_{P}}$ whose first component is
the unit $\bold{1}$. It is a subgroup of the set of invertible
elements. In a more general setting, there is a functor from the
category of augmented operads to the category of groups. For the
operad $P$, we have a commutative Hopf algebra structure on
$\mathbb{K}[\mathcal{G}_{P}]$ given by the set of coinvariants of
the operad.  It is a free commutative algebra of functions on
$\mathcal{G}_{P}$ generated by the set $(g_{\alpha})_{\alpha \in
A_{P}}$. Each $f \in \mathcal{G}_{P}$ can be represented by a formal
sum $f = \sum_{\alpha \in A_{P}} g_{\alpha}(f) \alpha$ such that
$g_{\bold{1}}=1$. There is a surjective morphism
\begin{equation} \label{f4}
\eta: g_{\alpha} \longmapsto \frac{F_{[\alpha]}}{|Aut(\alpha)|}
\end{equation}
from the Hopf algebra $\mathbb{K}[\mathcal{G}_{P}]$ to the incidence
Hopf algebra $H_{P}$ such that it shows that at the level of groups,
the affine group scheme $G_{P}(\mathbb{K})$ of $H_{P}$ is a subgroup
of the group $\mathcal{G}_{P}$. \cite{CL1, V1}

\begin{cor} \label{17}
(i) There is a surjective morphism from the Hopf algebra
$\mathbb{K}[\mathcal{G}_{NAP}]$ to the Connes-Kreimer Hopf algebra
$H_{CK}$ of rooted trees.

(ii) The affine group scheme $G_{CK}(\mathbb{K})$ of $H_{CK}$ is a
subgroup of $\mathcal{G}_{NAP}$.
\end{cor}

\begin{proof}
By \ref{8}, (\ref{f4}), $\rho \circ \eta:
\mathbb{K}[\mathcal{G}_{NAP}] \longrightarrow H_{CK}$ is a
surjective morphism of Hopf algebras and therefore in terms of
groups we will get the second claim.
\end{proof}

For the operad $NAP$, $\mathcal{G}_{NAP}$ is a group of formal power
series indexed by the set of unlabeled rooted trees and there is
also an explicit picture from the elements of $G_{NAP}(\mathbb{K})$
such that one can lift it to the universal Hopf algebra of
renormalization.

\begin{cor}  \label{18}
(i) Universal affine group scheme $\mathbb{U}(\mathbb{C})$ is a
subgroup of $\mathcal{G}_{NAP}$ and therefore each of its elements
will be representable by a formal power series indexed with Hall
rooted trees.

(ii) An element $F=\sum_{t} g_{t}(F)t$ of $\mathcal{G}_{NAP}$ is in
$\mathbb{U}(\mathbb{C})$ $\Longleftrightarrow$

- $t$ is a Hall tree in $\bold{H(T(A))}_{U}$, (It does not belong to
the Hopf ideal $I=ker \pi$.)

- And if $t=B^{+}_{f_{n}}(u)$ for some $u=t_{1}...t_{k}$ such that
$t_{1},...,t_{k} \in \bold{H(T(A))}_{U}$ and $f_{n} \in A$, one has
$$g_{t}(F)=\prod^{k}_{i=1} g_{B^{+}_{f_{n}}(t_{i})}(F).$$

\end{cor}

\begin{proof}
One can extend the morphism (\ref{f4}) to the level of the decorated
Hopf algebras and therefore by \ref{16} and \ref{17}, there is a
surjective map from $\mathbb{C}[\mathcal{G}_{NAP}](A)$ to $H_{U}$.
It shows that in terms of groups, $\mathbb{U}(\mathbb{C})$ is a
subgroup of $\mathcal{G}_{NAP}$. For the second case, according to
the lemma 6.12 in \cite{CL1}, each element of $\mathcal{G}_{NAP}$ is
in the subgroup $G_{NAP}$ if and only if for each tree
$t=B^{+}(t_{1}...t_{k})$, we have the following condition
$$ |sym(t)|g_{t}(F)=\prod^{k}_{i=1} |sym(B^{+}(t_{i}))|g_{B^{+}(t_{i})}(F).$$
By \ref{8}, \ref{9}, \ref{16}, (\ref{f4}), \ref{17} and since Hall
trees have no symmetry, the proof is complete.
\end{proof}

Let $H(A)$ be a Hall set and $A^{\star}:=\{ a^{\star}: a \in A \}$.
For each Hall word $w$, its associated {\it Hall polynomial} $p_{w}$
in the free Lie algebra $\mathcal{L}(A^{\star})$ is given as
follows:

- If $a \in A$, then $p_{a}=a^{\star}$,

- If $w$ is a Hall word of length $\ge 2$ such that its
corresponding Hall tree $t_{w}$ has the standard decomposition
$(t_{w_{1}},t_{w_{2}})$, then $p_{w}=[p_{w_{1}},p_{w_{2}}]$.

By induction, we can show that each $p_{w}$ is an homogeneous Lie
polynomial of degree equal to the length of $w$ and also it has the
same partial degree with respect to each letter as $w$.

In general, for each Hall set $H$, Hall polynomials form a basis for
the free Lie algebra (viewed as a vector space) and their decreasing
products $p_{h_{1}}...p_{h_{n}}$ such that $h_{i} \in H, \ \
h_{1}>h_{2}>...>h_{n}$, form a basis for the free associative
algebra (viewed as a vector space) \cite{R1}. About Hopf algebra
$H_{U}$, we identified a Hall set $\bold{H(T(A))}_{U}$ such that
$A=\{ f_{n}\}_{n \in \mathbb{N}_{>0}}$ and for each $f_{n}$ its
associated Hall polynomial is given by $p_{f_{n}}=e_{-n}$.

\begin{cor}  \label{19}
(i) As a vector space, Hall polynomials associated to the Hall set
$\bold{H(T(A))}_{U}$ form a basis for the Lie algebra
$L_{\mathbb{U}}$.

(ii) As a vector space, decreasing products of Hall polynomials with
respect to the associated Hall words to the Hall set
$\bold{H(T(A))}_{U}$ form a basis for the free algebra $H_{U}$.
\end{cor}

In the next section, we focus on the essential importance of $H_{U}$
in the mathematical reconstruction of perturbative renormalization
and with the help of its new picture (given by Hall rooted trees),
we are going to make a new representation from some important
physical information such as counterterms.

\section{Rooted tree representation of the universal singular frame}

The Riemann-Hilbert correspondence consists of describing a certain
category of equivalence classes of differential systems though a
representation theoretic datum. For a given renormalizable QFT
$\Phi$ with the related Hopf algebra $H$ and affine group scheme
$G$, one can identify a category of classes of flat equisingular
$G-$connections. It is a neutral Tannakian category and therefore it
should be equivalent with the category $\mathfrak{R}_{G^{\star}}$
(such that $G^{\star}:=G \rtimes \mathbb{G}_{m}$) of finite
dimensional linear representations of the affine group scheme of
automorphisms of the fiber functor of the main category. It is
reasonable to formulate the Riemann-Hilbert correspondence in a
universal setting by constructing the universal category
$\mathcal{E}$ of equivalence classes of all flat equisingular vector
bundles. This category can cover the corresponding categories
of all renormalizable theories and it means that when we are working
on the theory $\Phi$, we should consider the subcategory
$\mathcal{E}^{\Phi}$ of those flat equisingular vector bundles which
are equivalent to the finite dimensional linear representations of
$G^{\star}$. Since $\mathcal{E}$ is also a neutral Tannakian
category, it would be equivalent to a category of finite linear
representations of one special affine group scheme (related to the
universal Hopf algebra of renormalization), namely universal affine
group scheme $\mathbb{U}^{\star}$. In this case the fiber functor is
given by $\varphi: \mathcal{E} \longrightarrow
\mathcal{V}_{\mathbb{C}},$ $\Theta \longmapsto V$.  \cite{CM2, CM3,
CM1}

From this correspondence, one specific element will be determined
namely, the loop {\it universal singular frame}
$\gamma_{\mathbb{U}}$ with values in $\mathbb{U}$. Because of the
equivalence relation between loops (with values in the affine group
scheme $G$) and elements of the Lie algebra (corresponding to $G$)
(\cite{CM2, CM1}), for the presentation of the universal singular
frame, we should identify suitable element from the Lie algebra
$L_{\mathbb{U}}$ such that it is $e=\sum_{n \ge 1} e_{-n}$ (i.e. the
sum of the generators of the Lie algebra). Since the universal Hopf
algebra of renormalization is finite type, whenever we pair $e$ with
an element of the Hopf algebra, it would be only a finite sum and hence
$e$ does make sense. With help of the given rooted tree version of $H_{U}$ in previous part, we are going to make a new
representation from the universal singular frame based on rooted trees.

\begin{thm} \label{49}
(i) $e$ is an element of $L_{\mathbb{U}}$.

(ii) $e: H_{U} \longrightarrow \mathbb{K}[t]$ is a linear map. Its
affine group scheme level namely, $ \bold{rg}:
\mathbb{G}_{a}(\mathbb{C}) \longrightarrow \mathbb{U}(\mathbb{C})$
is a morphism that plays an essential role to obtain the
renormalization group.

(iii) The universal singular frame is given by
$\gamma_{\mathbb{U}}(z,v)=Te^{-\frac{1}{z} \int^{v}_{0} u^{Y}(e)
\frac{du}{u}}$.

(iv) For each loop $\gamma(z)$ in $Loop(G(\mathbb{C}),\mu)$, with
help of the associated representation $\rho: \mathbb{U}
\longrightarrow G$, the universal singular frame
$\gamma_{\mathbb{U}}$ maps to the negative part $\gamma_{-}(z)$ of
the Birkhoff decomposition of $\gamma(z)$ and also the
renormalization group $F_{t}$ in $G(\mathbb{C})$ is obtained by
$\rho \circ \bold{rg}$. \cite{CM2, CM3}
\end{thm}

For a given smooth manifold $M$, let $C(\mathbb{R}^{+})$ be the ring of
piecewise real valued continuous functions on $\mathbb{R}^{+}$ and
$\{X_{a}\}_{a \in A}$ be a family of smooth vector fields on $M$.
Suppose $\mathcal{A}$ be the algebra over $C(\mathbb{R}^{+})$ of
linear operators on $C^{\infty}(M)$ generated by the vector fields
$X_{a}$ ($a \in A$). For a family $\{g_{a}\}_{a \in A}$ of elements
in $C(\mathbb{R}^{+})$, set
\begin{equation} \label{f11}
X(x)=\sum_{a \in A} g_{a}(x)X_{a}.
\end{equation}
It can be expanded as a series of linear operators in $\mathcal{A}$
of the form $\sum_{w} g_{w}X_{w}$ such that

- $w=a_{1}...a_{m}$ is a word on $A$,

- $X_{1}=Id$ (identity operator), $X_{w}=X_{a_{1}}...X_{a_{m}}$,

- $g_{w} =  \int_{a_{m}} ... \int_{a_{1}} 1_{C(\mathbb{R}^{+})}$
where each $\int_{a_{i}}: C(\mathbb{R}^{+}) \longrightarrow
C(\mathbb{R}^{+}),$ $(1 \le i \le m)$ is a linear endomorphism
defined by

\begin{equation} \label{f12}
\{ \int_{a_{i}}g \}(x):= \int_{0}^{x} g(s)g_{a_{i}}(s)ds.
\end{equation}

Generally, for the given associative algebra $\mathcal{A}$ over the
commutative ring $\mathbb{K}$ generated by the elements $\{ E_{a}:a
\in A \}$, all elements in $\mathcal{A}$ are identified by formal
series $\sum_{w} \mu_{w}E_{w}$ such that $\mu_{w} \in \mathbb{K}$.
If $\mathcal{A}$ is a free algebra, then this representation will be
unique. For the Hall set $\bold{H(T(A))}$ of labeled rooted
trees with the corresponding Hall forest $\bold{H(F(A))}$, one can
assign elements $E(u)$ given by

- $E(\mathbb{I})=\mathfrak{e}$ (the unit of the Lie algebra
$\mathfrak{a}$ of $\mathcal{A}$),

- For each Hall tree $t$ with the standard decomposition
$(t^{1},t^{2}) \in \bold{H(T(A))} \times \bold{H(T(A))}$,
\begin{equation}
E(t)=[E(t^{2}),E(t^{1})]=E(t^{2})E(t^{1}) - E(t^{1})E(t^{2}),
\end{equation}

- For each $u \in \bold{H(F(A))} - \bold{H(T(A))}$ with the standard
decomposition $(u^{1},u^{2}) \in \bold{H(F(A))} \times
\bold{H(T(A))}$,
\begin{equation}
E(u)=E(u^{2})E(u^{1}).
\end{equation}

\begin{lem} \label{51}
(i) The Lie algebra $\mathfrak{a}$ is spanned by $\{ E(t): t \in
\bold{H(T(A))}\}$. It is called {\it Hall basis}.

(ii) $\mathcal{A}$ is spanned by $\{ E(u): u \in \bold{H(F(A))}\}$.
It is called {\it PBW basis}. \cite{R1}
\end{lem}

\begin{prop} \label{52}
For the locally finite total order set $\{f_{n}\}_{n \in
\mathbb{N}}$, the universal singular frame is represented by
$$\gamma_{U}(-z,v)= \sum_{n \ge 0, k_{j}>0} \alpha^{U}_{f_{k_{1}}f_{k_{2}}...f_{k_{n}}} p_{f_{k_{1}}}...p_{f_{k_{n}}} v^{\sum k_{j}} z^{-n} $$
such that $p_{f_{k_{j}}}$s are Hall polynomials.
\end{prop}

\begin{proof}
By \ref{49}, we know that
$$\gamma_{\mathbb{U}}(z,v)=Te^{-\frac{1}{z} \int^{v}_{0} u^{Y}(e) \frac{du}{u}}.$$
After the application of the time ordered exponential, we have

\begin{equation} \label{f15}
\gamma_{\mathbb{U}}(-z,v)= \sum_{n \ge 0, k_{j}>0}
\frac{e_{-k_{1}}...e_{-k_{n}}}{k_{1}(k_{1}+k_{2})...(k_{1}+...+k_{n})}
v^{\sum k_{j}} z^{-n}.
\end{equation}
In this expansion, the coefficient of the term
$e_{-k_{1}}...e_{-k_{n}}$ is given by the iterated integral
\begin{equation}
\int_{0 \le s_{1} \le ... \le s_{n} \le 1}
s_{1}^{k_{1}-1}...s_{n}^{k_{n}-1} ds_{1}...ds_{n}.
\end{equation}
By \ref{shuffle-order}, one can see $H_{U}$ as a free polynomial
algebra on the Lyndon words on the set $\{f_{n}\}_{n \in
\mathbb{N}_{>0}}$. Consider formal series
\begin{equation}
E:= f_{k_{1}}+x f_{k_{2}}+x^{2} f_{k_{3}}+...
\end{equation}
where
\begin{equation}
\mu_{k_{j}}(x)=x^{k_{j}-1}.
\end{equation}
By (\ref{f12}), for the variables $0 \le s_{1} \le ... \le s_{n} \le
1$, we have
\begin{equation}
\{\int_{k_{j}} 1\}(s_{j})= \int_{0}^{s_{j}} x^{k_{j}-1}dx.
\end{equation}
For each word $f_{k_{1}}f_{k_{2}}...f_{k_{n}}$, we can define the
following well-defined iterated integral
\begin{equation}
\alpha^{U}_{f_{k_{1}} f_{k_{2}}...f_{k_{n}} } := \int_{k_{n}} ...
\int_{k_{1}} 1.
\end{equation}
It is easy to see that the above integral is agree with the iterated
integral associated to the coefficient of the term
$e_{-k_{1}}...e_{-k_{n}}$. From the equations (\ref{order}) and
(\ref{f15}), the proof is complete.
\end{proof}

Proposition \ref{52} determines uniquely a real valued map on the
set $\bold{F(A)}$.

\begin{defn} \label{un}
Let $A=\{f_{n}\}_{n \in \mathbb{N}}$ be the locally finite total
order set corresponding to the universal Hopf algebra of
renormalization. For the given map in \ref{52} that associates to
each word $w=f_{k_{1}}f_{k_{2}}...f_{k_{n}}$ a real value
$\alpha^{U}_{w}$ and with notice to (\ref{f2}), define a new map
$\alpha^{U}$ on $\bold{F(A)}$ such that

- $\mathbb{I} \longmapsto \alpha^{U}(\mathbb{I})=1,$

- For each non-empty labeled forest $u$ in $\bold{F(A)}$,
$$ \alpha^{U}(u)= \sum_{>_{\mathfrak{u}(A)}} \alpha^{U}_{w(>_{\mathfrak{u}(A)})}. $$
\end{defn}
It is observed that for labeled rooted trees $t_{1},...,t_{m} \in
\bold{T(A)}$ and $f_{k_{j}} \in A$,
\begin{gather} \label{uns}
\alpha^{U}(t_{1}...t_{m})=\alpha^{U}(t_{1})...\alpha^{U}(t_{m}), \ \
\alpha^{U}(B^{+}_{f_{k_{j}}}(t_{1}...t_{m}))=\int_{k_{j}}
\alpha^{U}(t_{1})...\alpha^{U}(t_{m}),
\end{gather}
The map $\alpha^{U}$ (determined by the universal singular frame),
with the above properties, is uniquely characterized.

With notice to (\ref{f2}), for a given map $\alpha: \bold{F(A)}
\longrightarrow \mathbb{K}$, if $\alpha(\mathbb{I})=0$, then the
{\it exponential map} is defined by

- $exp \ \alpha(\mathbb{I})=1,$

- For each $u \in \bold{F(A)} - \{\mathbb{I}\}$,
\begin{equation}
exp \ \alpha(u)= \sum_{k=1}^{|u|} \frac{1}{k!} \alpha^{k}(u),
\end{equation}
and if $\alpha(\mathbb{I})=1$, then the {\it logarithm map} is
defined by

- $log \ \alpha(\mathbb{I})=0,$

- For each $u \in \bold{F(A)} - \{\mathbb{I}\}$,
\begin{equation}
log \ \alpha(u)= \sum_{k=1}^{|u|} \frac{(-1)^{k+1}}{k} (\alpha -
\epsilon)^{k}(u)
\end{equation}
such that $\epsilon(\mathbb{I})=1$ and for $u \in \bold{F(A)} -
\{\mathbb{I}\}$, $\epsilon(u)=0$. \cite{M1}

With help of the proposition 7 in \cite{M1}, for the defined map
\ref{un} with the properties (\ref{uns}), there exists a real valued
map $\beta^{U}$ on $\bold{F(A)}$ given by
\begin{equation} \label{betaa}
\alpha^{U}= exp \ \beta^{U}
\end{equation}
such that for each $u \in \bold{F(A)} - \bold{T(A)}$,
\begin{equation}
\beta^{U}(u)=0.
\end{equation}

\begin{prop} \label{53}
$$\sum_{k_{j}>0,n \ge 0} \alpha^{U}_{f_{k_{1}}f_{k_{2}}...f_{k_{n}}} f_{k_{1}}f_{k_{2}}...f_{k_{n}} =
exp \ (\sum_{t \in \bold{H(T(A))}_{U}} \beta^{U}(t)E(t))$$ such that

- $\{E(t): t \in \bold{H(T(A))}_{U}\}$ (the set of all Hall
polynomials) is the Hall basis for $L_{\mathbb{U}}$,

- $\{E(u): u \in \bold{H(F(A))}_{U}\}$ (the set of decreasing
products of Hall polynomials) is the PBW basis for $H_{U}$.

\end{prop}

\begin{proof}
It is observed that for a given map $\alpha: \bold{F(A)}
\longrightarrow \mathbb{K}$, if $\alpha(\mathbb{I})=0$, then
\begin{equation}
exp \ (\sum_{w} \alpha_{w}E_{w}) = \sum_{u \in \bold{H(F(A))}} exp \
\alpha(u)E(u),
\end{equation}
and if $\alpha(\mathbb{I})=1$, then
\begin{equation}
 log \ (\sum_{w} \alpha_{w}E_{w})= \sum_{u \in \bold{H(F(A))}} log \ \alpha(u)E(u).
\end{equation}
Now by (\ref{uns}) and with help of the continuous BCH formula (44)
in \cite{M1}, for each word $w=f_{k_{1}}f_{k_{2}}...f_{k_{n}}$ on
the set $A$, one can have
\begin{equation}
exp \ (\sum_{t \in \bold{H(T(A))}_{U}} \beta^{U}(t)E(t)) = \sum_{w}
\alpha^{U}_{w} w.
\end{equation}
Hall basis and PBW basis related to $H_{U}$ are determined with
notice to \ref{19} and \ref{51}.
\end{proof}

With notice to \ref{52}, \ref{un} and \ref{53}, a Hall rooted tree
representation from $\gamma_{\mathbb{U}}$ is obtained.

\begin{defn} \label{54}
The formal series $\sum_{t \in \bold{H(T(A))}_{U}} \beta^{U}(t)E(t)$
(given by \ref{53}) is called {\it Hall polynomial representation}
of the universal singular frame.
\end{defn}

There is also another way to show that the universal singular frame
has a rooted tree representation. \ref{16} and (\ref{f4}) provide a
surjective morphism from $\mathbb{C}[\mathcal{G}_{NAP}](A)$ to the
Hopf algebra $H_{U}$ and also corollary \ref{18} shows that the
universal affine group scheme $\mathbb{U(\mathbb{C})}$ is a subgroup
of $\mathcal{G}_{NAP}$. Since $\gamma_{\mathbb{U}}$ is a loop with
values in $\mathbb{U(\mathbb{C})}$, therefore for each fixed values
$z,v$, $\gamma_{\mathbb{U}}(z,v)$ should be a formal power series of
Hall rooted trees with the given conditions in \ref{18}.

For each loop in $Loop(G(\mathbb{C},\mu))$, the universal singular
frame maps to negative part of the Birkhoff decomposition of the
given loop such that it determines counterterms, renormalization
group and $\beta-$function of the theory. With notice to this fact
and with help of the introduced representation by \ref{49} (related
to the given loop), one can map the Hall polynomial representation
of the universal singular frame to these physical information and it
means that with the help of formal sums of Hall trees (Hall forests)
and Hall polynomials associated to the universal Hopf algebra of
renormalization, one can obtain new representations from these
physical information of the theory. This possibility is another
reason for the universality of $H_{U}$.

\end{document}